# Issues in Implementing Regression Calibration Analyses

Lillian Boe[1], Pamela A. Shaw[1,2], Douglas Midthune[3], Paul Gustafson[4], Victor Kipnis[3], Eunyoung Park[1], Daniela Sotres-Alvarez[5], Laurence Freedman[6,7] on behalf of the Measurement Error and Misclassification Topic Group (TG4) of the STRATOS Initiative

1. Department of Biostatistics, Epidemiology, and Informatics, University of Pennsylvania Perelman School of Medicine, Philadelphia, PA
2. Biostatistics Unit, Kaiser Permanente Washington Health Research Institute, Seattle, WA
3. Biometry Research Group, Division of Cancer Prevention, National Cancer Institute, Bethesda, MD
4. Department of Statistics, The University of British Columbia, Vancouver, BC, Canada
5. Department of Biostatistics and the Collaborative Studies Coordinating Center, Gillings School of Global Public Health, University of North Carolina, Chapel Hill, NC, USA
6. Biostatistics and Biomathematics Unit, Gertner Institute for Epidemiology and Health Policy Research, Sheba medical Center, Tel Hashomer, Israel
7. Information Management Services, Inc., Rockville, MD

**Abstract:**

Regression calibration is a popular approach for correcting biases in estimated regression parameters when exposure variables are measured with error. This approach involves building a calibration equation to estimate the value of the unknown true exposure given the error-prone measurement and other confounding covariates. The estimated, or calibrated, exposure is then substituted for the true exposure in the health outcome regression model. When used properly, regression calibration can greatly reduce the bias induced by exposure measurement error. Here, we first provide an overview of the statistical framework for regression calibration, specifically discussing how a special type of error, called Berkson error, arises in the estimated exposure. We then present practical issues to consider when applying regression calibration, including: (1) how to develop the calibration equation and which covariates to include; (2) valid ways to calculate standard errors (SE) of estimated regression coefficients; and (3) problems arising if one of the covariates in the calibration model is a mediator of the relationship between the exposure and outcome. Throughout the paper, we provide illustrative examples using data from the Hispanic Community Health Study/Study of Latinos (HCHS/SOL) and simulations. We conclude with recommendations for how to perform regression calibration.

## 1. Introduction

Measurement error in exposures that are important to public health, such as dietary intakes, physical activity and smoking, is becoming increasingly recognized as a problem [1-5], and methods for mitigating its effects are being researched and applied [6-11]. A recent review found regression calibration to be the most commonly used method to adjust for bias in estimates of an exposure-health outcome association resulting from measurement error [11]. However, implementing regression calibration requires care. Here, we highlight issues that commonly arise and illustrate them with examples drawn from the Hispanic Community Health Study/Study of Latinos (HCHS/SOL) [12] and simulations. This work is part of the STRengthening Analytical Thinking for Observational Studies (STRATOS) Initiative, an international effort aiming to provide accessible biostatistical guidance to elevate current practice in analyzing observational studies [13,14]. STRATOS comprises several working groups, including the Measurement Error and Misclassification Topic Group (TG4), focusing on areas of statistics where gaps exist between currently available statistical methodology and practice [13,15].

Regression calibration [6,16,17] involves developing a calibration equation that estimates the true exposure value for an individual based on their error-prone measured value and other covariates. This calibration estimate is then used in place of the unknown true exposure in the health outcome regression model. In Section 2, we explain the statistical basis for this approach, noting that the calibration estimate is itself an error-prone measure of the exposure, but one with a particular type of error, called Berkson error [18]. The method requires that this Berkson error be uncorrelated with the other outcome-model covariates.

We then consider three main topics:

1. The calibration equation – the validation study data required and choice of covariates. We emphasize the relation between the covariates in the health outcome model and those in the calibration model.

2. The outcome model and appropriate methods for computing the standard error (SE) of the estimated association parameter.

3. The special problem when one of the calibration-model covariates is a mediator of the relationship between exposure and health outcome.

We conclude with a check-list of the issues raised.

## 2. The Basics

### 2.1 *Regression Calibration*

We aim to learn the association between exposure X and health outcome Y, adjusted for confounding covariates Z. For instance, Y could be hypertension, X potassium intake, and Z age and sex. With data available on (Y,X,Z), fitting a regression model of Y on X and Z (the *outcome model*) allows examination of this association. We assume the model in question is a generalized linear model (including both linear and logistic regression as possibilities) or a Cox regression model, and that the correctly specified outcome regression model includes X as a linear effect with no interactions. Estimating the regression coefficient of X then accomplishes the task. We focus on X being a single exposure measure, but the concepts described can be extended to several exposures considered simultaneously [16].

When X is measured with error, as occurs when potassium intake is self-reported, more statistical effort is required. The available data are now (Y,X*,Z), where X* is an error-prone measurement of X. We assume the measurement error is *non-differential with respect to the outcome*, meaning that X* carries no more information about Y than is already provided by X and Z. In our example, this occurs if people with and without hypertension misreport similarly their potassium intake, a reasonable assumption if diet is reported before any diagnosis of hypertension. Even under this condition, plugging the (Y,X*,Z) data into the outcome model and estimating the regression coefficient of $X^*$ gives a biased estimate of the association of hypertension with potassium intake [3,5,8].

A popular method for avoiding this bias, *regression calibration,* is based on estimating each participant's unobserved X value given their observed $X^*$ and Z values. In our example, we estimate true potassium intake from self-reported intake and other covariates. The outcome model is now fit to (Y,$\hat{X}$,Z) data, where $\hat{X}$ is the said estimate. Underlying theory prescribes that this estimate be formed as the conditional mean of X given $X^*$ and Z, i.e., $\hat{X} = E(X|X^*,Z)$. This expression is called the *calibration equation*. We call the resulting value of $\hat{X}$ the *calibration estimate* of X. In some linear models, regression calibration exactly removes the bias, but more generally, this is only approximately true, including in our example where hypertension, Y, a binary variable, is modeled using logistic regression. For logistic and Cox regression models, the remaining bias is small when there is only a modest association between X and Y, a relatively small amount of measurement error in X*, or a rare outcome Y [6,17,19].

Using the calibration estimate $\hat{X}$ in the outcome model does lead to a higher variance of the estimated regression coefficient of X. In general, the lower is the correlation between $\hat{X}$ and X, the greater is the increase in variance of the estimated regression coefficient and the lower is the resultant statistical power to detect the association [20].

Usually the calibration equation, $\hat{X} = E(X|X^*,Z)$ is unknown, and must be estimated from *validation data*. These could be a sample of individuals for which (X,$X^*$,Z) are observed. The estimated regression model of X on $X^*$ and Z approximates the needed equation for $\hat{X}$, but the extra uncertainty involved in this estimation should be accounted for in the subsequent analysis (see Section 4).

Ideally, the validation data would consist of observations on (X, $X^*$, Z), in our example, exact potassium intake, self-reported potassium intake and age and sex. However, sometimes we might have only ($X^{**}$,$X^*$,Z), where $X^{**}$ is an unbiased measurement of X with random error, so that $E(X^{**}|X^*,Z) = E(X|X^*,Z)$. Then the estimation procedure is still valid. In our example, X** could be 24-hour urinary potassium, which is considered an unbiased measurement of potassium intake [3,8]. Sometimes, X* may itself be an unbiased measurement of X with random error, in which case replicate measures of X* provide sufficient validation data (see Appendix, Part 1).

2.2 *Berkson Error*
With most error-prone measurements X*, the error in the measurement is correlated with X*. In contrast, any surrogate measure $\hat{X}$ of an exposure X, is said to exhibit *Berkson error* if $X = \hat{X} + U$, where the random error U has mean 0 and is <u>independent of $\hat{X}$</u>. Berkson error occurs in occupational health studies, when individuals are assigned exposures equal to the mean exposure of their occupational subgroup. Berkson error is generally viewed as not causing bias in estimating associations, which is sometimes true. If error U is non-differential with respect to outcome Y, and is uncorrelated with all confounders Z, then using $\hat{X}$ in place of X in the outcome model does yield an unbiased estimate of the regression coefficient [21]. As explained below, this fact is the basis for the regression calibration method.

### 2.3 *Regression calibration through the Berkson error lens*

A natural question arises about why regression calibration works. When we cannot measure X exactly, but have an error-prone measurement X\*, we are told that substituting X\* for X in the regression of Y on X and Z, gives a biased estimate of the association of X with Y. Regression calibration tells us to use instead $\hat{X} = E(X|X^*,Z)$ in place of X. But $\hat{X}$ itself is an error-prone measurement of X, so how have we improved our situation? The answer is that $\hat{X}$ has Berkson error, whereas X\* (usually) does not. Moreover, by defining $\hat{X} = E(X|X^*,Z)$ the Berkson error term, U, is guaranteed to be uncorrelated with the confounding covariates Z.

Viewing regression calibration this way reinforces the hindrances placed upon the study investigator. We must estimate the exposure using $\hat{X} = E(X|X^*,Z)$; that is, we must use the same variables Z in the calibration equation as in the outcome model. This prohibits using a single all-purpose calibration equation for an exposure. In our example, estimating potassium intake as a function of self-reported potassium intake, age and sex is appropriate only for assessing the association between potassium intake and hypertension *given age and sex.* If we adjust the association for another confounder, such as socioeconomic status, then that confounder needs to be included in the calibration equation.

Suppose an additional variable $\tilde{Z}$, not included among the outcome model confounders Z, improves the estimation of X. Its inclusion in the calibration equation will increase the correlation of the resulting $\hat{X}$ with X, thereby increasing the power to detect the association of X with Y [22]. Theory allows use of $\tilde{Z}$ to help estimate X, *if it provides no extra information about the outcome beyond that provided by X and Z.* In our example, $\tilde{Z}$ might be an indicator of whether the self-report was made on a weekday or weekend, allowing adjustment of estimated potassium intake accordingly. However, if $\tilde{Z}$ provides extra information about the outcome, then the investigator needs to add $\tilde{Z}$ to the confounders Z already in the outcome model.

### 3. Forming the calibration equation

*3.1 The required data*

To form the calibration equation E(X|X*,Z), data are needed on X*, Z and either X or an unbiased measure of X, denoted by X** (see Section 2.1). Where possible, these data should be collected in an internal validation study, i.e., the participants should be a subgroup of the main study cohort. Internal validation is preferable for several reasons. First, covariates Z should include the confounders of the outcome model. With internal validation, these are naturally available. Second, the method of measuring X* should be the same in the validation study as in the main study.

Third, the calibration equation derived from the validation study should be *transportable* (i.e., applicable) to the main study data [6: Sections 2.2.4-2.2.5)]. With an internal validation study, this property will likely hold. For external validation studies, even with X* measured as in the main study, the equality of E(X|X*,Z) in the two studies holds only when the joint distributions of X and Z in the two studies coincide. In summary, with external validation studies, extra care in applying regression calibration is needed [23].

The design and sample size of validation studies are discussed in Section 4.2.

*3.2 The calibration equation*

Section 2 provides principles for constructing the calibration equation. Here we discuss implementation details and provide examples.

When the outcome model is specified using a transformed exposure (e.g. logarithmic scale), the calibration equation should be specified to estimate the transformed exposure directly [6: Section 4]. It is generally inappropriate to first estimate the untransformed exposure, and then transform the estimate. This applies equally when using a spline for modeling the exposure-outcome relationship (e.g., [24]). See Appendix, Part 2 for details.

The rule of including all outcome-model confounders Z in the calibration equation can be waived only when certain confounders demonstrably do not contribute to the estimation of X. In practice, one may test statistically whether confounders contribute to estimating X, and omit those that are seemingly unimportant. See Heinze et al [25], for example, for guidance on selecting covariates.

Including in the calibration equation an explanatory variable $\tilde{Z}$ not included in the outcome model has been termed *enhanced regression calibration* [22]. It is valid under the condition described in Section 2.3, which may be tested by examining whether $\tilde{Z}$ should or should not be selected for the outcome model [25].

Including $\bar{Z}$ in the calibration is useful only when it provides information about X over and above that provided by X* and Z [22]. In the example of Section 2.3, inclusion of the weekday/weekend report variable should contribute enough to the calibration model for log potassium intake to justify its selection [25].

To demonstrate the principle that the calibration and outcome models need to include the same confounders, we conducted simulations of a logistic regression outcome model with multivariate normal covariates X*, X, Z, and an additional confounder V. Error-prone exposure X* was correlated with X, Z and V. A validation subset included an unbiased biomarker X** with independent random error. Full details are provided in Appendix Part 3. For each simulation we fit the outcome model with X replaced by (i) the unadjusted X*, (ii) the calibration estimate $\bar{X}$ calculated from the correct calibration model, (iii) the calibration estimate $\hat{X}$ calculated from a non-aligned calibration model that omitted V from the covariates, and (iv) as in (ii) but with a non-aligned outcome model that omitted V from the covariates. The resulting estimated exposure coefficients are summarized in Table 1. Only the correctly performed regression calibration method (ii) yielded an (approximately) unbiased estimate.
[TABLE 1 HERE]

We illustrate the same principle with a real example. The HCHS/SOL is a large community-based cohort study of 16,415 Hispanic/Latinos in the US, recruited using a complex survey design [12]. For details, see Appendix Part 4. We examined the association between potassium intake and hypertension. Log potassium intake averaged over two 24-hour recalls was the error-prone [26] exposure measure (X*). We used HCHS/SOL internal validation sub-study, SOLNAS [26], data on 24-hour urinary excretion of potassium (X**), to develop a calibration equation for log potassium intake. The calibration equation included X* and all outcome model confounders, Z: age, sex, Hispanic/Latino background, education, income, current smoking, body mass index (BMI), and supplement use (yes/no). Using the calibration estimate $\hat{X}$ in the logistic regression outcome model, we estimated the odds ratio (OR) of hypertension for a 20% increase in potassium intake, controlling for confounders. We fit the outcome model first including all confounders, Z, and then including all except supplement use, to assess the impact of omitting a calibration equation covariate.
[TABLE 2 HERE]
With supplement use included in the outcome model, the estimated OR was 0.76 (95% CI: 0.60, 0.96), compared to 0.90 (95% CI: 0.75, 1.07) when omitted (see Table 2). Thus, incorrectly omitting this calibration-model covariate from the outcome model gave an OR much closer to 1.0 and no longer statistically significant.

## 4. The variance of the estimated exposure-outcome association

*4.1 Correctly estimating uncertainty of the association*

Because regression calibration uses a calibration estimate, $\widehat{X}$, of exposure, there is more uncertainty in the estimated exposure-outcome association compared to when using X. The extra uncertainty derives from two sources. The main source is the imperfect correlation between $\widehat{X}$ and X arising from the measurement error in X*. The second source is the finite sample available for estimating the calibration equation, leading to error in the estimated coefficients. Fitting the outcome model with $(Y,\widehat{X},Z)$, the SEs of the estimated outcome model coefficients reported by standard statistical software, such as *glm* in R or *genmod* in SAS, do not incorporate this second source of uncertainty, so are generally too small. This results in 95% confidence intervals (CI) being too narrow and having less than 95% coverage probability.

To demonstrate this problem, we used the simulation study described in Section 3. For correctly formulated regression calibration, we compared the SEs reported by *glm* in R with those based on a nonparametric bootstrap procedure accounting for the uncertainty in the calibration equation coefficients. We also compared the coverage of their 95% CIs.

[TABLE 3 HERE]

Table 3 shows that the model-based SEs (hereafter called "unadjusted") were too small, as judged by the empirical SE, and resulted in less than 95% coverage probability. The bootstrap-based SE was close to the empirical SE, providing close to 95% coverage.

In the HCHS/SOL example (Section 3), the unadjusted SE estimate of the regression coefficient for log potassium was 11% smaller than the SE estimate adjusted for calibration equation uncertainty (unadjusted SE = 0.59, corrected SE = 0.66). Consequently, the unadjusted 95% CI for the OR for a 20% increase in potassium intake was (0.61,0.94), compared to the adjusted 95% CI (0.60, 0.96). Here, the difference between the CIs was small; in other settings (e.g., with smaller validation studies) more appreciable differences are seen.

Two general methods to calculate a valid SE of regression calibration-based estimates are: (i) the bootstrap and (ii) "stacked estimation equations" methods. The bootstrap is usually easier to apply, with readily available statistical software, and widely applicable, but is computationally intensive; it requires fitting the calibration model and then the outcome model on many (generally ≥1000) bootstrap samples [27]. With an internal validation study, bootstrap sampling should be stratified by validation study participation. Example R code for implementing a bootstrap analysis is provided at the GitHub repository https://github.com/PamelaShaw/STRATOS-TG4-RC.

Details of the "stacked estimation equations" method [28] are provided in Appendix, Part 5. In some settings, the analytic formulas of Rosner et al [15,29,30] may be appropriate, and in studies with complex survey designs other methods are needed, as in our use of the multiple imputation variance estimator of Baldoni et al [31] for the HCHS/SOL study.

*4.2 Size and design of a validation study*

Uncertainty in the coefficients of the calibration equation increases the uncertainty in the estimated exposure-outcome association. One way of reducing uncertainty in estimating the calibration equation is to increase the validation study sample size. Keogh et al [21] provide guidelines for choosing the validation study size, including a sample size formula. Alternatively, simulations may be conducted, as in designing the SOLNAS validation study. See Appendix, Part 6.

Sampling design issues arise when planning internal validation studies. A simple random sample from the parent cohort is the simplest valid option. SOLNAS participants were recruited from the parent cohort using stratified sampling to ensure an equal distribution of participants from each of the four field centers. Consideration was given to adequate representation from each of the six Hispanic/Latino ethnicities (Cuban, Dominican, Mexican, Puerto Rican, Central American, and South American) and BMI categories. Increasing the sampling probabilities of individuals from small strata improves precision of calibration equation coefficients, while conditioning on the design variables in the calibration equation ensures valid estimates [32].

## 5. The dilemma of mediation of a covariate in the calibration equation

Covariates in the calibration model should usually be included as covariates in the outcome model (Section 2). However, a methodological problem can arise.

In outcome models, mediators (variables believed to lie on a causal pathway between exposure and outcome) should not be entered as covariates [33]. Doing so makes the exposure regression coefficient represent, not the total effect of exposure on outcome (as required), but only the non-mediated part of the effect. With regression calibration, this can create a dilemma. If a mediator variable is also important for estimating exposure, and is included in the calibration model, then it *should* be included in the outcome model to avoid biased estimation, but, as a mediator, the variable *should not* be included.

This dilemma was first encountered in nutritional epidemiology when studying the association of cancer with total energy intake [7]. Total energy intake is poorly reported using self-report instruments [34], but including BMI in the calibration model greatly improves the R-squared. In that case, BMI should be included in the outcome model. This creates two problems, one philosophic and one practical.

The philosophic problem is whether to regard BMI is a confounder or a mediator of the energy intake-disease association. Does high BMI cause higher energy intake, or does high energy intake cause higher BMI? If the former, then BMI is a confounder and should be included in the outcome model. If the latter, then BMI is a mediator and should be excluded.

Assuming BMI is a confounder allows its inclusion in the outcome model, but this still leads to a practical problem. BMI, being the most important covariate in the calibration equation, is highly correlated with calibrated energy, so entering both in the outcome model leads to collinearity and difficulty with estimating their separate associations with disease [9].

Assuming BMI is a mediator (which seems more likely), leads to the dilemma described above. In reference [35], Douglas Midthune, proposed a solution: BMI and calibrated energy intake are both included in the outcome model, and their coefficients are estimated and then combined linearly, accounting for the mediation and yielding an unbiased estimate. See Appendix, Part 7. Collinearity is partly overcome in this approach, because the linear combination of the two parameters can be estimated more precisely than each association separately, due to the large negative correlation between them.

We illustrate this method with another example from HCHS/SOL. We considered the association between energy intake and an outcome related to metabolic syndrome [36], adjusted for confounders, Z:age, sex, Hispanic/Latino background, education, income, and current smoking. Calibrated log energy intake, $\widehat{X}$, based on SOLNAS data, was a linear combination of the log average of two 24-hour recall energy intakes, confounders Z, and BMI. Note that we assume that participants' usual diet and weight were stable at the time of data collection.

We examined three methods of estimating the OR per 20% increase in energy intake.

1) Including BMI in the outcome model: i.e., regressing outcome on $\widehat{X}$, BMI, Z (assuming BMI is a confounder).

2) Omitting BMI from the outcome model: i.e., regressing outcome on $\widehat{X}$, Z (assuming BMI is a mediator – but known to give a biased estimate).

3) Midthune's method (assuming BMI is a mediator - see Appendix, Part 7).

(TABLE 4 HERE)

Table 4 shows that the three methods yielded widely different estimates. Including BMI in the model estimated an OR less than 1, with 95% CI including the null value. Omitting BMI from the outcome model gave a large, highly statistically significant OR (3.76), but one that is biased.

Midthune's method yielded a statistically significant, but much smaller OR (1.52), suggesting a positive total association between energy intake and the outcome. The 95% CIs for these latter two methods did not even overlap.

## 6. Discussion

After explaining the statistical principles involved, this manuscript considers practical issues arising when implementing regression calibration for addressing exposure error in epidemiological studies. To conclude, we provide a check-list of seven main points made in the earlier sections.

1. To avoid bias, *the calibration equation should include all confounders included in the outcome model*. Exceptions to this rule occur if a particular confounder does not contribute  to estimation of the exposure. This can be investigated while building the calibration equation using variable selection methods (e.g., see Heinze et al [25]). The above principle means that, f*or any given exposure, there is no single calibration equation that is appropriate for all analyses*.

2. *If at all possible, a validation study should be conducted internally*. The simplest valid design for an internal validation study is a simple random sample of participants in the main study. Sometimes more complex sampling designs can improve efficiency [37,38,39].

3. *The validation study should be large enough that the uncertainty in the calibration equation plays only a minor role in the precision of the estimated association.* A sample size formula is provided in Appendix, Part 6.

4. When constructing the calibration equation, *the equation's dependent variable should have the same functional form of the exposure that is used in the outcome model.* For example, if log exposure is used in the outcome model, the dependent variable of the calibration model should be log exposure.

5. To avoid bias, *the outcome model should include as covariates not only the relevant confounding variables, but also any extra covariates that are in the regression calibration model*. An exception is when there is evidence that the extra covariate is independent from the outcome, conditional on the other covariates in the model. This evidence may include showing that the covariate would not be selected for the outcome model [25].

6. *When regression calibration is used, SEs must be adjusted to account for the uncertainty*

*in the estimation of the calibration equation.* Approaches for SE estimation in the presence of regression calibration include the bootstrap and stacked estimating equations methods.

7. *When a calibration model covariate mediates the exposure-outcome relationship, Midthune's method of estimating the association parameter should be used.*

We reviewed common issues encountered when applying regression calibration and have made recommendations for how to address them. Regression calibration is an intuitive approach for addressing measurement error in covariates but its implementation requires care. When properly applied, it greatly reduces the bias in estimated association parameters caused by exposure measurement error [40]. Its wider use in observational epidemiology is recommended.

**Acknowledgements**

This work was supported in part by NIH grant R01-AI131771 (PS, LB) and by NHLBI contract 75N92019D00010 (DSA). HCHS/SOL study was carried out as a collaborative study supported by contracts from the National Heart, Lung, and Blood Institute (NHLBI) to the University of North Carolina (N01-HC65233), University of Miami (N01-HC65234), Albert Einstein College of Medicine (N01-HC65235), Northwestern University (N01-HC65236), and San Diego State University (N01-HC65237). This work was conducted on behalf of the STRATOS Measurement Error and Misclassification Topic Group (Topic Group 4). Membership of Topic Group 4 can be found at https://www.stratos-initiative.org/en/group_4.

**Conflicts of Interest:** None declared.

**Computer code:** R Code used for conducting the analyses described in this paper, together with simulated data similar to those used in the analyses, may be accessed at the website: https://github.com/PamelaShaw/STRATOS-TG4-RC.

Table 1: The mean estimate, empirical standard error of the mean, and mean percent (%) bias for 1000 simulated datasets of the log odds ratio (β) of X from uncorrected logistic regression, regression calibration (RC) correctly performed, RC with a non-aligned outcome model, and RC with a non-aligned calibration model. True β=log(1.5)=0.4055.

| Method | Mean of log odds ratio estimates | Empirical Standard Error of Mean[1] | % Bias |
|---|---|---|---|
| Uncorrected | 0.201 | 0.002 | -50.3 |
| Correct RC | 0.407 | 0.004 | 0.3 |
| RC, Non-aligned outcome model[2] | 0.912 | 0.006 | 125.0 |
| RC, Non-aligned calibration model[3] | 0.366 | 0.004 | -9.7 |

1. Empirical standard error of log odds ratio estimate/√(number of simulations)

2. Outcome model that omits the confounder V

3. Calibration model that omits the confounder V to produce a non-aligned calibration estimate $\overline{X1}$

Table 2: Data example from HCHS/SOL, presenting estimates of the odds ratio (OR) of hypertension associated with a 20% increase in potassium intake when supplement use is either included or excluded from outcome model.

| Method of Estimation | OR | 95% CI* |
|---|---|---|
| Including supplement use in outcome model | 0.76 | 0.60 – 0.96 |
| Omitting supplement use from outcome model | 0.90 | 0.75 – 1.07 |

* Based on a multiple imputation procedure described by Baldoni et al [31], with 25 imputations, see Appendix, Part 8.

Table 3: The mean estimate, mean percent (%) bias, empirical standard error, median estimated standard error, and coverage probabilities (CP) of the estimated 95% confidence interval for the log odds ratio (β) of X for regression calibration (RC) using uncorrected model-based estimates versus bootstrap-based estimates[a]. Based on 1000 simulated datasets. True β=log(1.5) =0.4055.

| Method | Mean of log odds ratio estimates | % Bias | Empirical standard error of log odds ratio estimate | Average estimated standard error | Coverage probability |
|---|---|---|---|---|---|
| Model-based | 0.407 | 0.3 | 0.136 | 0.113 | 0.915 |
| Bootstrap-based | | | | 0.140 | 0.954 |

[a] Bootstrap sampling was stratified on membership in the validation sub-study. We performed 1000 bootstrap iterations. Bootstrap confidence intervals are based on the percentile bootstrap.

Table 4: Data example from HCHS/SOL, presenting estimates of the odds ratio (OR) of at least one of four metabolic syndrome risk factors* associated with a 20% increase in energy intake, using three different methods

| Method of Estimation | OR | 95% CI** |
|---|---|---|
| Including BMI in outcome model | 0.85 | 0.47 – 1.53 |
| Omitting BMI from outcome model | 3.76 | 3.06 – 4.62 |
| Midthune's method | 1.52 | 1.01 – 2.27 |

\* Hypertension, Hyperlipidemia, Hypercholesterolemia, or Hyperglycemia
\** Method for 95% confidence intervals (CI's) is described in Appendix, Part 8.

**Supplementary Materials**

**Appendix**

**Part 1: Validation data when exposure measurement is unbiased for true exposure with random error**

When X* is unbiased for X and has random error independent of X and covariates Z, then the calibration equation E(X | X*, Z) can be estimated from validation data (X*$_1$, X*$_2$, Z), i.e. replicate measurements of X* together with covariates Z, and without having to determine the true exposure X. This is possible because E(X*$_1$ | X*$_2$, Z) = E(X | X*, Z), so that regressing X*$_1$ on X*$_2$ and Z will provide a valid calibration equation. In practice, there are more efficient ways of estimating the calibration equation from such data (see Section 4.4.2 of reference 6 in the main text).

**Part 2: Regression calibration when the exposure is modeled using a spline function**

Suppose the outcome Y is related to exposure X (or some chosen transformation of X) and confounders Z through a generalized linear model, which we write as:

$$E(Y|X, Z) = h\{\beta_0 + \beta_X f(X) + \beta_Z Z\} . \qquad (1)$$

Here $h$ is the inverse of the link function used in generalized linear models, allowing inclusion of both linear and logistic regression as possibilities. The function f is a transformation of the exposure variable. The object of our enquiry is to estimate the association parameter $\beta_X$.

As explained in Section 3 of the main text, because in the outcome model (1), we are using a transformed X, our calibration equation should now be defined as $\widehat{X} = E(f(X)|X^*, Z)$. In other words, one needs to estimate not the main exposure but its transformed value that is being used in the outcome model. This rule applies equally to the case of using a spline function for modeling the main exposure. For example, when using cubic restricted splines [main text reference 22], the outcome is modeled as a linear combination of two or more mathematical functions of X, $f_1(X)$, $f_2(X)$, etc. In this case, regression calibration equations should be formed for each of these separate functions of X, namely, $\widehat{X}_1 = E(f_1(X)|X^*, Z)$, $\widehat{X}_2 = E(f_2(X)|X^*, Z)$ and so on. The form of each regression equation could differ. Following this step, $\widehat{X}_1$, $\widehat{X}_2$, etc are used as explanatory variables with Z in the regression model for the outcome.

**Part 3: Computer simulations to demonstrate certain principles of regression calibration**

To demonstrate the principles that (a) the calibration equation needs to include confounders that are in the outcome model, and (b) the outcome model needs to include predictors that are in the calibration model, we conducted simulations. We generated 1000 simulations of a logistic regression model with a binary outcome Y dependent on multivariate normal covariates X, Z, and V for a cohort of N=2500 participants. We assumed that X* was observed instead of X, where X* was subject to the linear measurement error model $X^* = a_0 + a_1 X + a_2 Z + a_3 V + \varepsilon$. We also assumed that in a validation subset of n=250 participants, a biomarker X** with independent classical measurement error, i.e. $X^{**} = X + \delta$, with $E(\delta|X) = 0$, was observed. Values for all model parameters are provided in Table A1. For each simulated dataset we first regressed X** on X*, Z and V within the validation subset to obtain a "correct" calibration equation, and then applied that equation to the (X*,Z,V) values in the whole cohort so as to obtain the "correct" calibration estimate $\widehat{X}$ of X. We also regressed X** on just X* and Z (omitting V) within the validation subset to obtain a "non-aligned" calibration equation, and then applied that equation to the (X*,Z) values in the whole cohort so as to obtain a "non-aligned" calibration estimate $\widehat{X1}$ of X. We then fit (i) the logistic regression model with the uncorrected X* in place of X, and (ii) with the "correct" calibration estimate $\widehat{X}$ (calculated from the "correct" calibration model) in place of X, (iii) with the "non-aligned" calibration estimate $\widehat{X1}$ (calculated from the "non-aligned" calibration model) in place of X, and (iv) as in (ii) but with a non-aligned outcome model incorrectly omitting V from the confounders.

Table A1 Simulation Model and Parameters for

| Variables | Parameter values |
|---|---|
| $X^* = a_0 + a_1 X + a_2 Z + a_3 V + \varepsilon$ | $a_0 = 0.4$, $a_1 = 0.5$, $a_2 = 0.5$, $a_3 = 0.2$ ; $\sigma_\varepsilon^2 = 0.49$ |
| $X^{**} = X + \delta$ | $\sigma_\delta^2 = 0.7$ |
| (X,Z,V) | Multivariate normal $$\mu_X = \mu_Z = \mu_V = 0$$ $$\sigma_X^2 = \sigma_Z^2 = \sigma_V^2 = 1$$ cor(X,Z)= cor(X,V)=cor(Z,V) =0.5 |
| Logit(Y) = $b_0 + b_1 X + b_2 Z + b_3 V$ | $b_0 = -1.0$, $b_1 = \log(1.5)$, $b_2 = -\log(1.3)$, $b_3 = \log(1.75)$ |

**Part 4: The Hispanic Community Health Study/ Study of Latinos (HCHS/SOL)**

HCHS/SOL [main text reference 12] is a large community-based cohort study of Hispanic/Latinos in the US (n=16,415). The HCHS/SOL study consists of self-identified Hispanic/Latino adults of Cuban, Puerto Rican, Dominican, Mexican, and Central and South American backgrounds. Participants were recruited from four United States field centers (Chicago, Illinois; Miami, Florida; Bronx, New York; San Diego, California) using a complex survey-design, and the baseline clinic visit (2008-11) included anthropometry and blood pressure measures, a fasting blood sample, health questionnaires, and two 24-hour dietary recalls. The study includes an internal validation study called the Study of Latinos Nutritional and Physical Activity Assessment Study (SOLNAS) (n=485) which includes a subset of participants, in addition to the above measures, the recovery biomarkers doubly-labeled water for energy intake, and 24-hour urinary nitrogen, potassium and sodium for intakes of protein, potassium and sodium [main text reference 24].

## Part 5: Stacking estimation equations for calculating the standard error of regression calibration estimates of outcome model regression coefficients

The "stacked estimation equations" method for estimating the sandwich standard error is described by Stefanski and Boos [main text reference 28]. Carroll et al [main text reference 6 (Appendix Section A.6.6)] provide further details of its use in the specific setting of regression calibration. This approach has the advantage of being less computationally intensive (only one sample is analyzed), but it requires the practitioner to do more specialized calculations in statistical software. However, use of numerical derivatives can often simplify the work, and the method can be straightforward to implement once it is learnt. An example can be found in Shaw et al [reference A1], who applied a special extension of regression calibration in which both the outcome and covariate were estimated. These authors provided explicit expressions for the stacked estimating equations in their Appendix.

## Part 6: Sample size for validation studies used for yielding calibration equations

We provide here a guideline for choosing the appropriate size of a validation study that is conducted to yield a calibration equation. We restrict discussion to an internal validation study where the participants may be viewed as a simple random sample of the participants in the main study. A first principle is that the size of the validation study is decided in relation to its contribution to the main study's goal. Therefore, we must specify, among other things, what is the desired statistical power for testing the exposure-disease association in the main study.

We assume a single exposure X that is measured by X* with non-differential error, and where the relationship between X and X* is linear. We focus on the aim of estimating the slope in a simple linear regression model of a health outcome Y as a function of X, i.e. we consider the model:
$E(Y|X) = \beta_0 + \beta_X X$.
If we were to use X* in place of X, then we would obtain instead a different regression model:
$E(Y|X^*) = \beta_0^* + \beta_X^* X^*$.

With these simple models, the relation $\beta_X* = \lambda\,\beta_X$ holds, where $\lambda$ is the attenuation coefficient, defined as the coefficient of X* in the linear regression of X on X*. Therefore, a simple "adjusted estimate" of $\beta_X$ is obtained by dividing the estimate of $\beta_X*$ (that comes from the main study) by an estimate of $\lambda$ that can be obtained from the validation study.

Applying the Delta method, the variance of this adjusted estimate of $\beta_X$ can then be expressed by the following approximation [main text reference 29]:

$$var(\hat{\beta}_X) = \frac{var(\widehat{\beta_X^*})}{\hat{\lambda}^2} + \frac{(\widehat{\beta_X^*})^2 var(\hat{\lambda})}{\hat{\lambda}^4}$$

(The "hats" placed above the symbols denote estimated quantities.) This formula assumes the estimates of $\lambda$ and $\beta_X$ are independent, which would be exactly true if the validation study were external to the main study, and approximately true if the validation study were internal and its sample size was only a small fraction of the main study sample size, as is usually the case.

The second term on the right-hand side of the above expression represents the extra uncertainty introduced into the estimate of $\beta_X$ by the uncertainty in the value of $\lambda$. We may choose the size of the validation study to minimize the impact of this extra uncertainty, specifically so that the second term of the right-hand side will be a small fraction, $f$, of the first term. This leads to the equation:

$$var(\hat{\lambda}) = \frac{f\hat{\lambda}^2 var(\widehat{\beta_X^*})}{(\widehat{\beta_X^*})^2}$$

Since $var(\hat{\lambda})$ is dependent on the sample size of the validation study, this equation allows us to determine the size of that study. In fact, it leads to the following equation.

Suppose the investigator is planning to test, in the main study, the association of disease with exposure at a two-sided significance level $\alpha$ with power $1-\omega$. Then the required sample size of the validation study, $n_v$, is:

$$n_v = \frac{\left\{\Phi^{-1}\left(1 - \frac{\alpha}{2}\right) + \Phi^{-1}(1 - \omega)\right\}^2 (1 - \rho_{X^*X}^2)}{f\rho_{X^*X}^2}$$

where $\rho_{X^*X}^2$ is the squared correlation coefficient between X and X*, and where $\Phi^{-1}$ is the inverse of the standard normal distribution function.

For example, if $f = 0.1$, $\alpha = 0.05$, $1-\omega = 0.90$ and $\rho_{X^*X} = 0.4$, then a sample size $n_v = 552$ will ensure that the variance of the adjusted estimate of $\beta_X$ will not increase by more than about 10% ($f = 0.1$) as a result of the uncertainty in estimating $\lambda$. Clearly, the larger the degree of measurement error, the smaller will be $\rho_{X^*X}$, the correlation between X and X*, and the larger the validation/calibration study that will be needed. Also, the higher the power $(1 - \omega)$ desired in the main study, the larger will be the required validation study.

Note that the quantity $\rho_{X^*X}$ will not be known precisely before the validation study is conducted—indeed it is one of the quantities that we hope to estimate from the validation study data. Consequently, a plausible value will need to be specified in order to use the above sample size formula. This is not unlike needing to specify an exposure's hitherto unknown effect on the outcome in order to calculate the sample size of a main study.

The above formula applies when the validation study includes, as the reference measure, and exact measurement of X. However, if the reference measurement is X**, an unbiased but inexact measurement of X, then the formula needs to be modified by using, in place of $\rho_{X^*X}$ (the correlation between X and X*), the correlation between X** and X*. This will tend to be smaller than $\rho_{X^*X}$ due to the measurement error in X**, and will thus further increase the required sample size of the validation study.

As an alternative to using a sample size formula, for the SOLNAS internal validation study, simulations were conducted to show that a sample size of approximately 500 was sufficient to provide at least 80% power to detect a hazard ratio for diabetes of 1.8 for a 30% increase in energy intake (Investigator communication, Shaw). These simulations were conducted using a method of correction for regression calibration in Cox regression and were based on the approaches presented in Shaw and Prentice [reference A2].

## Part 7: Regression calibration for mediation models

### Part 7.1 Mediation models:

Let     Y = outcome variable

         X = exposure of interest

         Z = confounder(s)

         M = mediator

The goal of mediation analysis is to explore potential pathways that could help explain an observed relationship between exposure X and outcome Y. The analysis assumes a causal effect of X on Y (X → Y) and hypothesizes that at least part of the effect can be explained by X's effect on mediator M (X → M → Y). The mediation model can be expressed as two regressions, one regressing M on X and Z, the other regressing Y on X, Z and M. When both regressions are linear, the model can be written as

$$M = \gamma_0 + \gamma_X X + \gamma_Z Z + \delta, \tag{A1}$$

$$Y = \beta_0 + \beta_X X + \beta_Z Z + \beta_M M + \varepsilon, \tag{A2}$$

where random variables $\delta$ and $\varepsilon$ have zero means and are independent of each other and of X and Z. Substituting the right-hand side of equation (A1) for M in equation (A2), we get

$$Y = \tilde{\beta}_0 + \tilde{\beta}_X X + \tilde{\beta}_Z Z + \tilde{\varepsilon}, \tag{A3}$$

where $\tilde{\beta}_X = \beta_X + \beta_M \gamma_X$, $\tilde{\beta}_Z = \beta_Z + \beta_M \gamma_Z$, $\tilde{\beta}_0 = \beta_0 + \beta_M \gamma_0$, and $\tilde{\varepsilon} = \varepsilon + \beta_M \delta$ has mean zero and is independent of X and Z. In mediation analysis, regression coefficient $\beta_X$ is called the 'direct' effect of X on Y, $\beta_M \gamma_X$ is called the 'indirect' effect (mediated by M), and $\tilde{\beta}_X$ is called the 'total' effect (adjusted for confounder Z).

In our application, Y is a binary variable and logistic regression is used to model the probability that Y = 1 (p). For logistic regression, equations (A2) and (A3) are replaced by

$$\text{logit}(p) = \beta_0 + \beta_X X + \beta_Z Z + \beta_M M, \tag{A2a}$$

$$\text{logit}(p) \approx \tilde{\beta}_0 + \tilde{\beta}_X X + \tilde{\beta}_Z Z , \tag{A3a}$$

where $\text{logit}(p) = \log\{p / (1 - p)\}$, and the decomposition of total effect into direct and indirect effects, $\tilde{\beta}_X \approx \beta_X + \beta_M \gamma_X$, is only approximate, since mediation analysis typically relies on direct and indirect effects defined in terms of linear regression model coefficients. Thus, for a binary outcome, we rely on the approximate linearity of the logistic model.

## Part 7.2: Regression calibration to estimate total effect $\tilde{\beta}_X$

In most situations, researchers want to estimate total effects ($\tilde{\beta}_X$) rather than direct or indirect effects, which is why the 'mediator principle' says to include confounders but not mediators in a regression model. Direct application of the mediator principle is problematic, however, when trying to correct for bias due to measurement error in X. In this appendix, we show how to use mediation models to correct for bias in estimated total effects due to measurement error in X.

Suppose we have a large main study with variables (Y, X*, Z, M), where X* is a surrogate of X, and a smaller validation study with variables ($X_1^{**}$, $X_2^{**}$, X*, Z, M), where $X_1^{**}$ and $X_2^{**}$ are repeat measurements of a reference measure that is unbiased for X. We assume the following measurement error models for X* and $X_k^{**}$, k = 1, 2,

$$X^* = \alpha_0 + \alpha_X X + \alpha_Z Z + \alpha_M M + e^*, \tag{A4}$$

$$X_k^{**} = X + e_k^{**},$$

where random errors $e^*$, $e_1^{**}$ and $e_2^{**}$ have zero means and are independent of each other and of X, Z and M. We also assume that $X^*$ has non-differential error with respect to model (A2), i.e., that $X^*$ and Y are conditionally independent given X, Z and M.

If we applied regression calibration directly to outcome model (A3), we would fit calibration model $\widehat{X} = E(X \mid X^*, Z)$ and substitute $\widehat{X}$ for X in (A3). The problem with this approach is that regression calibration requires measurement error to be non-differential *with respect to the outcome model*, and even when we assume $X^*$ has non-differential error with respect to model (A2), it will still have differential error with respect to model (3) unless $\alpha_M = 0$ or $\beta_M = 0$.

Similarly, for 'expanded' regression calibration (including M in the calibration model but not the outcome model) to be unbiased, Y and M must be conditionally independent given X and Z (or X and M must be conditionally independent given Z), which again is not the case if M is a mediator.

Instead, we estimate total effects by applying regression calibration separately to each component of the mediation model (A1)-(A2):

Step 1: Fit calibration model $\widehat{X}^{(1)} = E(X_2^{**} \mid X_1^{**}, Z)$ in the validation study.

Step 2: Substitute $\widehat{X}^{(1)}$ for X in model (A1) to estimate $\gamma_X$ in the validation study.

Step 3: Fit calibration model $\widehat{X}^{(2)} = E(X_1^{**} \mid X^*, Z, M)$ in the validation study.

Step 4: Substitute $\widehat{X}^{(2)}$ for X in model (A2) to estimate $\beta_X$ and $\beta_M$ in the main study.

Step 5: Estimate $\tilde{\beta}_X$ using the equation $\tilde{\beta}_X = \beta_X + \beta_M \gamma_X$.

As noted previously, the equation in Step 5 is exact for linear regression but only approximate for logistic regression. Step 1 requires the validation study to have repeat measurements of the reference instrument on at least a subset of participants, although an alternative estimate of $\gamma_X$ could be obtained with only one measurement per participant if the variance of $e_1^{**}$ were known from other sources. The above 5 steps constitute Midthune's method of estimating the total effect of X on Y.

**Part 7.3: Potential bias applying regression calibration to model (A3)**

We illustrate the potential bias in $\tilde{\beta}_X$, the total effect of X, when regression calibration is applied directly to outcome model (3), in the special case where there are no confounders Z. We show $\tilde{\beta}_X$ is biased whether or not the mediator M is included in the calibration model.

**Standard regression calibration:** With no confounders in outcome model (A3), the standard regression calibration model is

$$\widehat{X}^{(s)} = E(X \mid X^*) = \lambda_0 + \lambda_X X^*,$$

where $\lambda_X = \text{cov}(X, X^*)/\text{var}(X^*)$. The standard regression calibration estimator derived from substituting $\widehat{X}^{(s)}$ for X in (A3) has asymptotic mean

$$\begin{aligned}
\widetilde{\beta}_X^{(s)} &= \text{cov}(Y, \widehat{X}^{(s)})/\text{var}(\widehat{X}^{(s)}) = \text{cov}(Y, X^*)/\text{cov}(X, X^*) \\
&= \beta_X + \beta_M \text{cov}(M, X^*)/\text{cov}(X, X^*) \\
&= \beta_X + \beta_M(\alpha_X \sigma_{XM} + \alpha_M \sigma_M^2)/(\alpha_X \sigma_X^2 + \alpha_M \sigma_{XM}),
\end{aligned}$$

where $\sigma_X^2$ and $\sigma_M^2$ are the respective variances of X and M and $\sigma_{XM}$ is their covariance. The asymptotic bias of the standard regression calibration estimator is therefore

$$\text{bias}^{(s)} = \widetilde{\beta}_X^{(s)} - \widetilde{\beta}_X = \beta_M \alpha_M \sigma_M^2 (1 - \rho_{XM}^2)/(\alpha_X \sigma_X^2 + \alpha_M \sigma_{XM}),$$

where $\rho_{XM} = \sigma_{XM}/\sigma_X \sigma_M$ is the correlation of X and M. The estimator is therefore biased unless $\beta_M = 0$ or $\alpha_M = 0$ (or $\rho_{XM}^2 = 1$).

**Expanded regression calibration:** With no confounders in outcome model (A3), the expanded regression calibration model is

$$\widehat{X}^{(e)} = E(X \mid X^*, M) = \lambda_0 + \lambda_X X^* + \lambda_M M,$$

where $\lambda_X = \text{cov}(X, X^* \mid M)/\text{var}(X^* \mid M)$. The expanded regression calibration estimator derived from substituting $\widehat{X}^{(e)}$ for X in (A3) has asymptotic mean

$$\begin{aligned}
\widetilde{\beta}_X^{(e)} &= \text{cov}(Y, \widehat{X}^{(e)})/\text{var}(\widehat{X}^{(e)}) = \{\beta_X \text{cov}(X, \widehat{X}^{(e)}) + \beta_M \text{cov}(M, \widehat{X}^{(e)})\}/\text{var}(\widehat{X}^{(e)}) \\
&= \beta_X + \beta_M \text{cov}(M, X)/\text{var}(\widehat{X}^{(e)}) \\
&= \beta_X + \beta_M \gamma_X / R_{X\widehat{X}^{(e)}}^2,
\end{aligned}$$

where $R_{X\widehat{X}^{(e)}}^2 = \text{var}(\widehat{X}^{(e)})/\text{var}(X)$ is the proportion of the variance in X explained by estimator $\widehat{X}^{(e)}$. The asymptotic bias of the expanded regression calibration estimator is therefore

$$\text{bias}^{(e)} = \widetilde{\beta}_X^{(e)} - \widetilde{\beta}_X = \beta_M \gamma_X (1 - R_{X\widehat{X}^{(e)}}^2)/R_{X\widehat{X}^{(e)}}^2.$$

The estimator is therefore biased unless $\beta_M = 0$ or $\gamma_X = 0$ (or $R_{X\widehat{X}^{(e)}}^2 = 1$).

## Part 8: Variance Estimation and Confidence Limits of the Regression Calibration Estimate in the Outcome Model

We describe how we obtained the variance estimate for the odds ratio (OR) estimates presented in the data analysis in Section 5 of the main manuscript, namely for the OR of a 20% increase in energy intake in the multivariable logistic model for the high risk outcome. Variance estimation requires multiple steps to (1) account for the extra uncertainty added by the nuisance parameters estimated for the calibration and mediation models using the SOLNAS subset ([main text reference 24]; see also references in main text) and (2) account for the complex survey design of the HCHS/SOL, which involved clustering, stratification, and unequal probability sampling [reference A3; main text reference 12].

To address the extra uncertainty added by estimating the calibration and mediation model parameters, we used a variance estimate that relies on a resampling-based multiple imputation procedure, an approach first proposed by Baldoni et al [main text reference 31] for regression calibration to address covariate measurement error in the complex survey setting. In this approach, the expected value of the latent true exposure (usual dietary intake of potassium) was multiply imputed using two calibration models $\hat{X}_1^{(m)} = E(X|X^{**},Z)$ and $\hat{X}_2^{(m)} = E(X|X^*, Z,BMI)$ refit on a bootstrap sample of the SOLNAS subset, where X* is the error-prone self-reported measurement (i.e., potassium intake from a 24hr recall) and X** is the measurement with only random error (e.g., 24-hour urinary potassium excretion). Within each imputation step, the outcome and mediation models were fit with these calibrated values, as detailed in Appendix A Part 1, to derive the measurement error adjusted estimate of the total effect of energy and its bootstrap variance, $\hat{\bar{\beta}}^{(m)}$ and $\hat{V}^{(m)}$. The results of M imputations were then pooled to produce the variance of $\hat{\bar{\beta}}$ using the law of total variance. Detailed steps are described below.

First, at the m$^{th}$ imputation step, a new regression calibration model was fit on a bootstrap resample of the SOLNAS participants, where the bootstrap resample is stratified on membership to the reliability subset, for which there is a second biomarker observation approximately six months later. On the m$^{th}$ bootstrap replicate sample, we regressed the biomarker replicate $X_2^{**}$ on $X_1^{**}$ and other confounders, Z, in order to obtain estimates of the calibration model coefficients and to form the estimator $\hat{X}_1^{(m)} = E(X|X^{**},Z)$ for each individual in the SOLNAS sample. Similarly, for each bootstrap sample, we regressed the biomarker X** on the 24-hour recall energy intake measure, X*, BMI, and other confounders, Z, to obtain new estimates of the model coefficients for the 24-hour recall calibration model and a new calibrated intake $\hat{X}_2^{(m)} = E(X|X^*, Z)$ for the entire HCHS cohort. The resampling process was repeated M times so that there were M sets of (1) calibration coefficients for the two calibration models and (2) corresponding estimated intake values, $\hat{X}_1^{(m)}$ and $\hat{X}_2^{(m)}$ for the mediation and outcome models, respectively.

For the within imputation estimate of the total effect of energy $\hat{\bar{\beta}}^{(m)}$ and its variance $\hat{V}^{(m)}$, we accounted for the complex survey design of the HCHS/SOL study using a survey bootstrap procedure. HCHS/SOL bootstrap weights were calculated by first sampling primary sampling units (PSUs) with replacement and equal probability from each stratum. To ensure efficiency of the bootstrap estimators and rectify potential biases due to the small number of PSUs per stratum, the selected participants from each PSU had their base weights scaled, according to the recommendations from Kolenikov [reference A4]. Then, the scaled bootstrap weights were further adjusted by their respective response rates. Using the set of replicate weights, $w^{(r)}$, $r = 1, \dots, R$, we recalculated Midthune's estimate $\hat{\tilde{\beta}}^{(m,r)}$ R+1 times, where the first run used the original weights and all subsequent runs used the replicate weights to obtain the parameter estimates of interest. Specifically, the logistic regression of the high risk outcome on $\hat{X}_2^{(m)}$, BMI, and Z was fit using data from the full HCHS cohort and accounting for the complex survey design for each set of weights $w^{(r)}$, to obtain (R+1) estimated coefficients of calibrated log energy $\hat{\beta}_e^{(m,r)}$ and BMI $\hat{\beta}_{bmi}^{(m,r)}$. Similarly, the regression of BMI on $\hat{X}_1^{(m)}$ and Z was performed on a bootstrapped SOLNAS subset to produce a total of R estimated coefficients of biomarker calibrated log energy $\hat{\gamma}_e^{(m,r)}$. The within imputation estimate for the total effect of energy was then estimated as $\hat{\tilde{\beta}}^{(m)} = \hat{\beta}_e^{(m,1)} + \hat{\gamma}_e^{(m,1)}\hat{\beta}_{bmi}^{(m,1)}$, the $m^{th}$ estimate using the original survey weights. The corresponding bootstrapped variance estimate was $\hat{V}^{(m)} = \frac{1}{R-1}\sum_{r=1}^{R}\left(\hat{\tilde{\beta}}^{(m,r)} - \bar{\tilde{\beta}}^{(m)}\right)^2$, where $\hat{\tilde{\beta}}^{(m,r)} = \hat{\beta}_e^{(m,r)} + \hat{\gamma}_e^{(m,r)}\hat{\beta}_{bmi}^{(m,r)}$ and $\bar{\tilde{\beta}}^{(m)} = \frac{1}{R-1}\sum_{r=1}^{R}\hat{\tilde{\beta}}^{(m,r)}$.

The process of obtaining Midthune's estimate using the original and replicate weights was repeated M times, once for each of the newly estimated intake values $\hat{X}_1^{(m)}$ and $\hat{X}_2^{(m)}$ obtained from the multiple imputation process described above. For our analysis, we chose M=25 imputations and R=1000 bootstrap replicate weights. Following Baldoni et al. (2021), the variance estimate $\hat{V}$ of $\hat{\tilde{\beta}}$ was computed as $\hat{V} = \frac{1}{M}\sum_{m=1}^{M}\hat{V}^{(m)} + \frac{1}{M-1}\sum_{m=1}^{M}\left(\hat{\tilde{\beta}}^{(m)} - \hat{\tilde{\beta}}\right)^2$. To avoid any numerical instability, we used robust estimators for the mean and standard deviation (i.e. the median and median absolute deviation, respectively) in the equation above to calculate $\hat{V}$, as recommended by Baldoni et al [main reference 31].

**Appendix References**